\documentclass[12pt]{article}
\usepackage{amsfonts,amsmath,amssymb,natbib}

\title{Determinate Values for Quantum Observables}
\author{Roderich Tumulka}
\date{March 28, 2006}

\newcommand{\RRR}{\mathbb{R}}
\newcommand{\CCC}{\mathbb{C}}
\newcommand{\PPP}{\mathbb{P}}

\newcommand{\D}{\mathrm{d}} % differential d
\newcommand{\I}{\mathrm{i}} % imaginary i

\bibpunct{(}{)}{;}{a}{}{,}

\begin{document}
\maketitle
\begin{abstract}
This is a comment on J.~A.~Barrett's article ``The Preferred-Basis Problem and the Quantum Mechanics of Everything'' in Brit.\ J.\ Phil.\ Sci.\ \textbf{56} (2005), which concerns theories postulating that certain quantum observables have determinate values, corresponding to additional (often called ``hidden'') variables. I point out that it is far from clear, for most observables, what such a postulate is supposed to mean, unless the postulated additional variable is related to a clear ontology in space-time, such as particle world lines, string world sheets, or fields.

\medskip

\end{abstract}

In his recent article \citep{Bar05}, Jeffrey A.~Barrett developed an astute analysis of the problems that would arise for Bohmian mechanics if mental states did not supervene on the positions of the particles constituting the brain. My comment on his article is not so much a criticism but rather concerns a point that I think should be kept in mind in this context but that Barrett did not mention in his article. The point is that one is not always free to add a postulate like ``there exist determinate values for the quantum-mechanical observable $\Lambda$'' because it is far from clear, for most observables, what such a postulate would \emph{mean}. It is the purpose of this comment to elucidate this point. Related considerations can be found in \citep{DDGZ96}, in particular Section 6, and in \citep{AGTZ06}, in particular Sections 4--8.

According to Bohmian mechanics (also known as de Broglie--Bohm pilot-wave theory) \citep{Bohm52,Bell66,Gol01} there are point particles moving in 3-space, which have a determinate position at every time, in such a way that the configuration $q$ is random with probability distribution $|\psi_t(q)|^2$ at every time $t$. In the language of quantum mechanics, one could say that the position observables have determinate values in Bohmian mechanics, namely the actual positions of the particles, while other observables are not attributed determinate values (except in special cases, such as that of the wave function being an eigenfunction of the observable). Similarly, \citet{Bohm52} in his quantum field theory model postulated determinate values of the electromagnetic field operators, and \citet{Bell86} in his quantum field theory model postulated determinate values of the fermion number operators. Since one can understand in all of these cases what it means for these observables to have determinate values, one might get the impression that one can always postulate the existence of such additional (``hidden'') variables that provide the actual values for any observable $\Lambda$ one may wish to choose. However, that means to overlook the following problem.

While it is clear what it means to postulate that there are point particles moving in 3-space, it is not clear at all what it should mean to postulate that there is a variable $\lambda$, governed by certain equations, until we specify the relation of $\lambda$ to the physical ontology in space-time, such as particle world lines or string world sheets or fields. The relation could be that $\lambda$ is a function of the ontology in space-time, or that $\lambda$ influences its behavior; e.g., the velocity is a function of the particle world line, and the mass affects the world line as a parameter in the equation of motion (in Bohmian or classical mechanics).

Let me approach the problem by a sequence of examples. Suppose first that some theory, some alternative to Bohmian mechanics, postulates that not the positions, but the velocities are the observables that have determinate values.\footnote{The reader might think at this point that such a theory would not solve the measurement problem; I agree, but this is not the point I want to make now. The reader might also think that the way Bohmian mechanics ensures that the probability distribution is $|\psi_t(q)|^2$ at all times $t$, cannot be transferred to the momentum representation; again, I agree, but this is not the point I want to make now.} Then this postulate could be understood as saying that 
\begin{equation}\label{statem2}
\begin{split}
  &\text{there are point particles moving in 3-space whose velocities}\\
  &\text{$v_i$ 
  have joint probability distribution $|\tilde\psi_t(\ldots m_iv_i \ldots)|^2$}
\end{split}
\end{equation}
where $\tilde\psi$ is the momentum representation (Fourier transform) of $\psi$ and $m_i$ the mass of particle $i$. The point is that I did not simply say, 
\begin{equation}\label{statem1}
\begin{split}
  &\text{there are variables $v_i$ with joint probability} \\
  &\text{distribution $|\tilde\psi_t(\ldots m_iv_i \ldots)|^2$.}
\end{split}
\end{equation}
I would not be able to understand the meaning of \eqref{statem1} without further information, whereas \eqref{statem2} is clear. As an example of the ambiguity of \eqref{statem1}, I could as well postulate
\begin{equation}\label{statem3}
\begin{split}
  &\text{there are point particles moving in 3-space whose \textbf{positions}}\\
  &\text{$v_i$
  have joint probability distribution $|\tilde\psi_t(\ldots m_iv_i \ldots)|^2$.}
\end{split}
\end{equation}
This would seem compatible with \eqref{statem1}, indeed as compatible as \eqref{statem2}.\footnote{The reader might think that such a theory would not make the same predictions as quantum mechanics; I agree, but this is not the point I want to make now.} This fact indicates that \eqref{statem1} is an incomplete description of a theory and needs to be complemented by either \eqref{statem2}, or \eqref{statem3}, or a similar statement.

Suppose now, as a second example, that some theory postulates that 
\begin{equation}\label{statem6}
  \text{``energy is the observable with determinate values.''}
\end{equation}
In analogy to the previous example, in particular to \eqref{statem2}, this could be understood as saying that
\begin{equation}\label{statem4}
\begin{split}
  &\text{there are point particles moving in 3-space whose}\\
  &\text{positions $q_i$ and velocities $v_i$ have such a distribution }\PPP\\ 
  &\text{that 
  $\PPP\Bigl(\sum_i\tfrac{1}{2} m_i v^2_i + 
  V(\ldots q_i \ldots) = E\Bigr) = \langle \psi | P_E | \psi \rangle$}
\end{split}
\end{equation}
with $V$ the potential and $P_E$ the projection to the eigenspace of the Hamiltonian with eigenvalue $E$. But it could also be understood in a different way, for example as saying that
\begin{equation}\label{statem5}\begin{split}
  &\text{there is a field $\phi(x,t)$ on space-time whose distribution }\PPP\\
  &\text{is such that }\PPP\Bigl( \int \D^3x\, |\nabla\phi(x,t)|^2 = E\Bigr) = 
  \langle \psi |P_E | \psi\rangle.
\end{split}\end{equation}
Moreover, \eqref{statem4} or \eqref{statem5} is not what comes to mind when one reads \eqref{statem6}. What comes to mind, instead, is that there are no point particles (and no strings and no fields), but there is only \emph{energy as such}. And this is the step where it is not clear, at least to me, what the statement is supposed to mean.

The difficulty is connected to the question, What is the meaning of the word ``energy''? In classical mechanics or field theory, energy is defined to be $\sum_i\tfrac{1}{2} m_i v^2_i + V(\ldots q_i \ldots)$ respectively $\int \D^3x\, |\nabla\phi(x,t)|^2$. If I am supposed to understand the word ``energy'' without any further definition, I am lost. For example, I wonder what would then be the difference if the same determinate value $\lambda$ were not \emph{energy as such}, but, say, \emph{charge as such}? Moreover, part of the physicist's psychological expectation connected with the statement that ``energy has the value $\lambda$'' is that a measurement of the energy will have result $\lambda$. But \emph{that} cannot be postulated, but either follows or does not follow from the macroscopic motion of matter constituting the measurement apparatus; and this matter should correspond to some ontology in space-time.

As a third example, let us consider various ways of dealing with spin in theories like Bohmian mechanics. The modern way of treating spin, which goes back to \citet{Bell66}, postulates no further additional variable for spin in addition to the particle positions; spin is taken into account by assuming that the wave function of an electron is $\CCC^2$-valued, $\psi(q) = \bigl(\psi_{\uparrow}(q), \psi_{\downarrow}(q) \bigr)$, and the probability distribution of its position is $\|\psi(q)\|^2 = |\psi_{\uparrow}(q)|^2 + |\psi_{\downarrow}(q)|^2$ with $\|\cdot \|$ the norm in $\CCC^2$. Another possible approach that has been investigated by \citet{BH} and \citet{holland} assumes that an electron is a tiny rotating sphere, and that the spin observables are somehow connected to the actual speed of rotation. While there are reasons to prefer Bell's approach, the relevant aspect here is that it is understandable what ``speed of rotation'' means if an electron is an extended body with an orientation in space.\footnote{A side remark concerning a subtlety: There is an ambiguity about what it means to be a sphere. One possibility, providing what is needed to make sense of the notion of ``speed of rotation,'' is that every point on the sphere has a trajectory, identifying points on the sphere at different times. The other possibility is to postulate that the sphere is simply a subset of space filled with matter, without any identification of points at different times. In the latter view, the speed of rotation is not defined.} Let me contrast this with a postulate that is not understandable, exactly for the reason that it introduces a variable $\lambda$ without relating it to the constituents of the physical ontology. The fact that the wave function of an electron can be written as a function $\psi(q,s)$ of a continuous (position) variable $q \in \RRR^3$ and a discrete (spin) variable $s \in \{\uparrow, \downarrow\}$ may suggest to introduce, in addition to the particle position, another additional variable $\lambda \in \{\uparrow, \downarrow\}$ representing the ``actual ($z$-component of the) spin.''\footnote{The reader might think at this point that such a postulate would unduly prefer the $z$-axis over other directions; I agree, but this is not the point I want to make now.} My objection is that \emph{it does not make sense} to merely postulate ``there is a variable $\lambda$ which can assume the values $\uparrow$ and $\downarrow$ and obeys these-and-these laws.'' To make it meaningful, one might postulate that an electron is not a point particle but, say, a tiny (asymmetric) rod that always points in either the $z$ direction or the $-z$ direction, and that $\lambda$ indicates this direction; or one might specify how $\lambda$ influences the particle's trajectory in space.

Many readers may have the following question at this point: Would it not be sufficient for a newly introduced variable $\lambda$ if there was a psycho-physical law according to which my conscious experience is a certain function of $\lambda$? Would it not be unnecessary then to relate $\lambda$ to some physical ontology in space and time? To be sure, since such a theory would predict the same experience, it is irrefutable; but it has the flavor of a brain-in-the-vat scenario. According to such a theory, the physical world would not at all be like what we usually think it is, and that is why I do not seriously consider this possibility. The world would not contain any trees, or rocks, or chairs, as it would not be able to contain different amounts of matter in different regions of space and time. Instead, there would just be the abstract quantity $\lambda$, giving rise to illusory experiences of trees and rocks and chairs. I cannot stop believing that there are trees and rocks and chairs.

I hope that these examples help convey why I think one should regard the postulate ``the observable $\Lambda$ has determinate value $\lambda$'' as incomprehensible in a physical theory unless $\lambda$ is connected to a physical ontology in space-time such as particle world lines or string world sheets or fields. Such a connection is naturally suggested for some observables together with certain ontologies, such as position operators with particle ontology or field operators with field ontology. For the velocity operator $-\tfrac{\I}{m} \nabla$, a connection is less strongly suggested, and for the energy operator $H$ hardly at all. For most operators no connection suggests itself (a concrete example is $\Lambda = q-\I\nabla$).

For Barrett's \citeyearpar{Bar05} discussion this means that ``$E$-theory'' (according to which all quantum-mechanical observables have determinate values) appears not to be among the viable options, i.e., among the meaningful theories, as for most observables $\Lambda$ the postulate that $\Lambda$ has determinate values $\lambda$ remains incomprehensible. Whether ``$Q$-theory'' (according to which the observable $Q$ on which mental states supervene, whichever it may turn out to be, has determinate values) is a viable option seems to depend on which observable $Q$ turns out to be. For Bohmian mechanics, our discussion means that it is misleading to say, ``the position observables have determinate values.'' A more appropriate description of Bohmian mechanics is, ``there are particles moving in 3-space whose positions have the distribution of the quantum-mechanical position observables.''

\section*{Acknowledgments} 

I thank J.~A.~Barrett, S.~Goldstein, T.~Maudlin, and N.~Zangh\`\i\ for helpful discussions of the subject, and an anonymous referee for his helpful comments on an earlier version of this article.

\bigskip

\noindent
    Mathematisches Institut\\
    Eberhard-Karls-Unversit\"at\\
    Auf der Morgenstelle 10\\
    72076 T\"ubingen\\
    Germany\\
    tumulka@everest.mathematik.uni-tuebingen.de

\end{document}